# Design-controlled Synthesis of IrO$_2$ sub-monolayers on Au Nanodendrites: Marrying Plasmonic and Electrocatalytic Properties

Isabel C. de Freitas,[a] Luanna S. Parreira,[a] Eduardo C. M. Barbosa,[a] Barbara A. Novaes,[a] Tong Mou,[b] Tiago. V. Alves,[c] Jhon Quiroz,[d] Yi-Chi Wang,[e] Thomas J Slater,[e,f] Andrew Thomas,[e] Bin Wang,[b] Sarah J Haigh,[e] and Pedro H. C. Camargo[a,d]*

[a]*Departamento de Química Fundamental, Instituto de Química, Universidade de São Paulo, Avenida Prof. Lineu Prestes, 748, 05508-000 São Paulo, SP, Brazil*

[b]*Center for Interfacial Reaction Engineering and School of Chemical, Biological, and Materials Engineering, Gallogly College of Engineering, The University of Oklahoma, Norman, Oklahoma 73019, United States*

[c]*Departamento de Físico-Química, Instituto de Química, Universidade Federal da Bahia Rua Barão de Jeremoabo, 147, 40170-115, Salvador, BA, Brazil*

[d]*Department of Chemistry, University of Helsinki, A.I. Virtasen aukio 1, Helsinki, Finland*

[e]*School of Materials, University of Manchester, Manchester M13 9PL, United Kingdom*

[f]*Electron Physical Sciences Imaging Centre, Diamond Light Source Ltd., Oxfordshire OX11 0DE, United Kingdom*

*Corresponding author: Email: pedro.camargo@helsinki.fi

**Abstract**

We develop herein plasmonic-catalytic Au-IrO$_2$ nanostructures with a morphology optimized for efficient light harvesting and catalytic surface area; the nanoparticles have a dendritic morphology, with closely spaced Au branches all partially covered by an ultrathin (1 nm) IrO$_2$ shell. This nanoparticle architecture optimizes optical features due to the interactions of closely spaced plasmonic branches forming electromagnetic hot spots, and the ultra-thin IrO$_2$ layer maximizes efficient use of this expensive catalyst. This concept was evaluated towards the enhancement of the electrocatalytic performances towards the oxygen evolution reaction (OER) as a model transformation. The OER can play a central role in meeting future energy demands but the performance of conventional electrocatalysts in this reaction is limited by the sluggish OER kinetics. We demonstrate an improvement of the OER performance for one of the most active OER catalysts, IrO$_2$, by harvesting plasmonic effects from visible light illumination in multimetallic nanoparticles. We find that the OER activity for the Au-IrO$_2$ nanodendrites can be improved under LSPR excitation, matching best properties reported in the literature. Our simulations and electrocatalytic data demonstrate that the enhancement in OER activities can be attributed to an electronic interaction between Au and IrO$_2$ and to the activation of Ir-O bonds by LSPR excited hot holes, leading to a change in the reaction mechanism (rate-determinant step) under visible light illumination.

**Introduction**

Plasmonic catalysis relies on harvesting the energy generated by localized surface plasmon resonant (LSPR) excitations in plasmonic nanoparticles to drive, accelerate, and/or control molecular transformations.[1–6] Following LSPR excitation in plasmonic nanoparticles, non-radiative plasmon decay can lead to the formation of LSPR-excited charge carriers.[7,8] These LSPR-excited hot electrons and holes can electronically or vibrationally excite molecular adsorbates at the metal-molecule interface via direct or indirect mechanisms.[9,10] This can lead not only to improved reaction rates relative to the reaction in the absence of LSPR excitation, but also provide new reaction pathways for the control over reaction selectivity relative to traditional thermochemically-driven processes.[11–15] Gold (Au) and silver (Ag) nanoparticles are amongst the strongest plasmonic structures, supporting LSPR excitation in the visible and near-infrared ranges with wavelengths that are tunable via the control of shape, size, composition and structure.[16,17] Consequently plasmonic catalysis has emerged as an attractive approach for solar to chemical energy conversion,[1,3,18–22] with Au and Ag nanoparticles having been applied as plasmonic catalysts towards a variety of molecular transformations under visible-light excitation.[20,23–27]

Among several important chemical transformations, the water splitting reaction to produce hydrogen ($H_2$) and oxygen ($O_2$) has attracted massive attention for energy conversion and storage applications.[28–31] Unfortunately, this reaction is limited by significant efficiency loss and high overpotentials ($\eta$) as a result of the sluggish kinetics for the oxygen evolution reaction (OER, which represents the oxidative half-reaction).[32,33] It has been established that iridium and ruthenium oxides ($IrO_2$ and $RuO_2$, respectively) are among the best OER electrocatalysts, and $IrO_2$ is often used in proton exchange membrane water electrolyzers as a result of its higher durability relative to $RuO_2$.[34,35] However, to meet our future energy demands there is an urgent need to develop materials displaying improved OER electrocatalytic activites. For example, iridium and ruthenium are expensive so bulk oxide materials do not make the most efficient use of the material. Furthermore, $IrO_2$ and $RuO_2$, both require relatively high overpotentials and a reduction in the overpotential, and therefore an improvement in the energy efficiency of the OER is highly desirable.[36,37]

In this context, we believe that the harvesting of plasmonic effects represents an attractive strategy for the improvement of OER performances.[38–40] This approach has the potential to enable the use of solar light as an abundant and sustainable energy input to enhance OER rates. In fact, several plasmonic materials have been employed to enhance OER, hydrogen evolution reaction, and oxygen reduction reaction rates under light illumination.[41–45] Unfortunately one of the best OER materials, $RuO_2$ and $IrO_2$, do not support LSPR excitation in the visible or near infrared ranges.[46] However, the range of materials that support LSPR excitation in the visible or near-infrared ranges is limited to Ag, Au, Cu, Al, and Ni.[8,47–50] In order to bridge the gap between materials with the desired catalytic properties and LSPR materials, the synthesis of multimetallic nanoparticle architectures that enable one to combine catalytic and plasmonic components (and thus catalytic and plasmonic properties) has emerged as an effective approach.[11,51,52] In these plasmonic-catalytic nanoparticles, the goal is to use the plasmonic metal to harvest energy from light excitation, so that the generated LSPR charge-carriers can be transferred or dissipated to the surface of the catalytic material, where it can be further utilized to perform plasmon-driven chemistry.[8,53]

Inspired by this approach, we describe herein the development of a plasmonic-catalytic core-shell multimetallic nanostructure composed of Au and $IrO_2$ as the plasmonic and catalytic components, respectively. More specifically, the synthesized Au-$IrO_2$ plasmonic-catalytic nanoparticles displayed a tortuous dendritic morphology, in which several branches are closely spaced to each other and each branch is composed of Au partially covered by an ultrathin (1 nm) Ir-based shell. These features are very attractive to address the OER for a variety of reasons: *i*) $IrO_2$ represents one of the most active species towards the OER;[54,55] *ii*) the ultrathin and incomplete $IrO_2$ shell at the surface of each branch maximizes the light harvesting by Au and the subsequent flow of charge carriers from Au to the Ir-based shell;[56] *iii*) the ultrathin and incomplete Ir-based shell maximises the $IrO_2$ surface area and therefore minimizes the loading and utilization of this expensive metal; and *iv*) the high curvature and plasmonic hybridization between the closely spaced plasmonic branches allows for the generation of a high density of electromagnetic hot spots, *i.e.*, areas of high electric field enhancements at the junctions of the branches as a result of the LSPR excitation that can be felt by the catalytic Ir-based shells (as shown in **Figure S1A-C**).[11,51] These high electric field enhancements, for instance, can decay via absorption (**Figure 1D-F**) and lead to the formation of energetic charge carriers (hot electrons and holes) at the $IrO_2$, as described in antenna-reactor complex nanoparticle designs.[12,51,57,58] The success of this approach is demonstrated by the significant improvement in the OER activity for the multimetallic Au-$IrO_2$ nanodendrites due to LSPR excitation with visible light. The effect of the light excitation wavelength in determining the OER activity supports the role of the LSPR excitation in this transformation. Our data demonstrates that electronic/charge transfer interaction between Au and IrO$_2$ and the activation of Ir-O bonds by the LSPR excited hot holes contributes to the catalysts improved performance under visible light irradiation.

**Results and discussion**

Plasmonic-catalytic nanostructures have been synthesized in which the plasmonic and catalytic components are in direct contact and where the two components are separated by a small distance (~up to 5 nm).[12,51,52,59,60] Both these scenarios can allow for the acceleration of reaction rates and control over reaction selectivity over the surface of the catalytic metal[8,11,61,62] but the two architectures have different benefits.[52] In nanoparticle designs in which plasmonic and catalytic components are not in direct contact, it has been demonstrated that the catalytic component may be exposed to regions of local electric fields induced by the LSPR excitation of the plasmonic metal, enhancing catalytic activity.[10,12,57,63] In contrast with nanostructures where the catalytic and plasmonic components are in direct contact, electronic effects and charge flow from the plasmonic to the catalytic component occurs upon LSPR excitation.[11,51,58] Inspired by these recent findings, we have developed nanoparticle architectures that take advantage of both scenarios, *i.e.*, expose the catalytic component to regions of high electric-fields promoted by the neighboring plasmonic metal, as well as allowing the flow of energetic charge carriers from the plasmonic to the catalytic component as a result of LSPR excitation. Specifically, we find highly effective plasmonic-catalytic nanostructures are produced using Au as the plasmonic component, IrO$_2$ as the catalytic component, and adopting a complex core-shell nanodendrite morphology.

The synthesis of the multimetallic Au-IrO$_2$ nanodendrites was performed by the co-reduction of Au and Ir precursors (AuCl$_4^-$ and IrCl$_3 \bullet$xH$_2$O, respectively) in the presence of sodium citrate as both reducing agent and stabilizer.[56] **Figure 1A-E** shows scanning electron microscope (SEM) images (**Figure 1A**), high-angle annular dark field (HAADF) scanning transmission electron microscope (STEM) images (**Figure 1B** and **D**), and STEM energy dispersive X-ray (EDX) elemental maps (**Figure 1C** and **E**) for the Au-IrO$_2$ nanodendrites obtained by this approach. Here, the molar ratio between the AuCl$_4^-$ and IrCl$_3 \bullet$xH$_2$O precursors employed during the synthesis corresponded to AuCl$_4^-$:IrCl$_3 \bullet$xH$_2$O 1:1.5. It can be observed from the SEM images (**Figure 1A**) that the nanodendrites displayed an overall spherical morphology with a highly tortuous branched surface structure. The nanodendrite particles have an overall average diameter corresponding to 93.2 ± 9.0 nm and thus a monodisperse size distribution. The branched morphology is even more apparent in the HAADF-STEM images shown in **Figure 1B** and **D**. These images indicate that the dendrite branches are closely spaced and each branch has an approximately spherical cross section with a diameter of approximately 5 nm. HRTEM imaging confirms the nanodendrites are polycrystalline with all the observed lattice spacings being assigned to *fcc* Au (**Figure S2)**. This is further confirmed by the electron diffraction pattern from an individual nanodendrite as shown in **Figure S3**.

The STEM-EDX elemental mapping shown in **Figure 1C** and **E** indicates that the nanodendrite is composed of a core-shell morphology with a Au core covered by an ultrathin (1 nm or thinner) Ir rich surface layer. Closer inspection of the high-resolution elemental map (**Figure 1E**) demonstrates that the Ir-based ultrathin layer does not completely cover the surfaces of the Au branches (**Figure S4** shows an additional STEM-EDX high resolution elemental map where regions of exposed Au surface are highlighted). The overall nanoparticle composition from the STEM-EDX compositional maps in **Figure 1C** corresponded to Au 87 at% and Ir 13 at% which is in good agreement with the value obtained from flame atomic absorption spectroscopy (FAAS) (Au 86 at% and Ir 14 at%).

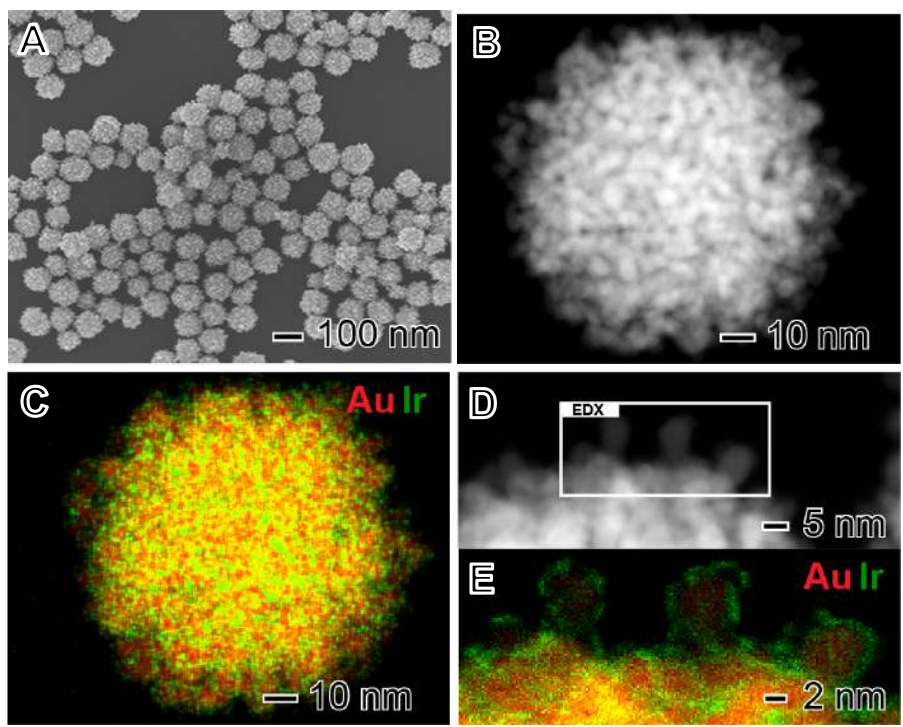

**Figure 1.** SEM (A), HAADF-STEM (D and D) and STEM-EDX (C and E) images of Au-IrO$_2$ nanodendrites. The nanodendrites were obtained by co-reduction of AuCl$_4^-$ and IrCl$_3$·xH$_2$O precursors in a 1:1.5 molar ratio. The Au and Ir at% in the samples corresponded to 85 and 15, respectively. The elemental distributions for Au and Ir are shown in red and green, respectively, in the STEM-EDX maps. Where both green and red signals are overlapping the colour map appears yellow

In order to investigate the effect of the molar ratio of AuCl$_4^-$ and IrCl$_3$•xH$_2$O precursors used during the synthesis on the morphological and compositional features of the nanodendrites, we varied the precursor molar ratios in the co-precipitation synthesis to AuCl$_4^-$:IrCl$_3$•xH$_2$O 1:0.25 (previously 1:1.5). **Figure 2A-E** shows the SEM (**Figure 2A**) and HAADF-STEM (**Figure 2B** and **D**) images and STEM-EDX elemental maps (**Figure 2C** and **E**) for the resulting nanodendrites. A similar nanodendritic morphology as described in **Figure 1** is produced**;** nanodendrites with an overall spherical shape, a monodisperse size distribution, closely spaced branches, and with each branch comprised of Au partially covered by an ultrathin and incomplete Ir-rich shell. However, two important differences were detected. As expected the nanodendrites contained less Ir, with an elemental composition of Au 96 at% and Ir 4 at% (data extracted from the STEM-EDX in **Figure 2C** and which agrees with the values obtained from FAAS for these samples). In addition, the individual branches were more elongated, having a rod-like morphology, relative to the nanodendrites shown in **Figure 1**. This more-elongated, rod-like morphology for the branches can be also visualized by additional HAADF-STEM images and STEM-EDX elemental maps shown in **Figure S5**.

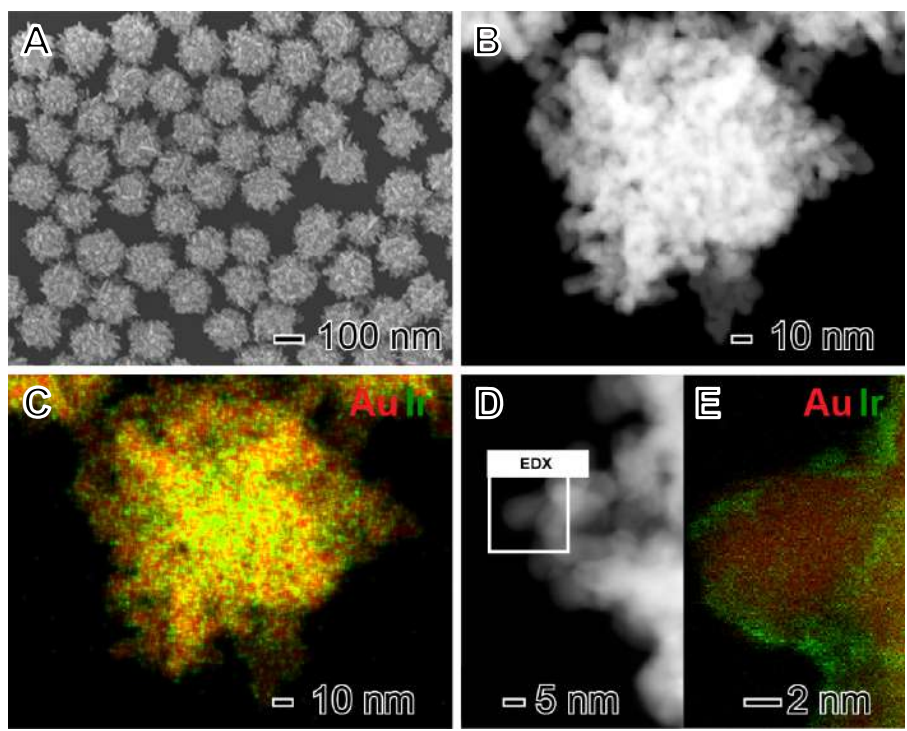

**Figure 2.** SEM (A), HAADF-STEM (C and D) and STEM-EDX (C and E) images for Au-IrO$_2$ nanodendrites obtained by co-reduction of AuCl$_4^-$ and IrCl$_3$·xH$_2$O precursors in a 1:0.25 molar

ratio. The Au and Ir at% in the samples corresponded to 96 and 4, respectively. The elemental distributions for Au and Ir are shown in red and green, respectively, in the STEM-EDX maps. Where both green and red signals are overlapping the colour map appears yellow.

The XRD diffractograms obtained for the nanodendrites prepared employing $AuCl_4^-$:$IrCl_3 \bullet xH_2O$ 1:1.5 and 1:0.25 molar ratios are shown in **Figure 3A** (red and black traces, respectively). The results agree with the compositional and morphological variations in the samples. With the decrease in the $IrCl_3 \bullet xH_2O$ precursor content employed in the synthesis, a decrease in the Ir content led to an increase in the intensity of the reflections assigned to *fcc* Au. Moreover, the XRD peaks from the *fcc* Au became less broad, consistent with the increased length of the dendrite branches observed by STEM, which will increase the size of the Au crystallites. No peaks assigned to Ir or $IrO_2$ phases could be detected in the samples, which is not unexpected given the ultrathin thickness of the Ir-based surface layer. Moreover, no XRD peaks at all were detected for a control sample prepared under identical conditions but in the absence of $AuCl_4^-$ precursor (blue trace).

**Figure 3B** depicts the UV-VIS extinction spectra recorded from aqueous suspensions containing the nanodendrites. It can be observed that both samples prepared employing $AuCl_4^-$:$IrCl_3 \bullet xH_2O$ 1:1.5 and 1:0.25 molar ratios (red and black traces, respectively) displayed an extinction peak in the visible-range centered at 576 nm. A decrease in the extinction intensity for the sample with higher Ir content ($AuCl_4^-$:$IrCl_3 \bullet xH_2O$ 1:1.5) was observed. This agrees with the fact that the larger Ir concentration in the samples prepared employing $AuCl_4^-$:$IrCl_3 \bullet xH_2O$ 1:1.5 can lead to a stronger suppression of the LSPR excitation (absorption and scattering) from the plasmonic (Au) component relative to the sample prepared employing $AuCl_4^-$:$IrCl_3 \bullet xH_2O$ 1:0.25 molar ratios (lower Ir content).[64] Au NPs (obtained in the absence of $IrCl_3 \bullet xH_2O$ precursor) displayed the characteristic LSPR dipolar band at 527 nm (**Figure S6A**),[16] while $IrO_2$ obtained in the absence of $AuCl_4^-$ precursor displayed no bands in the visible region (**Figure S6B**).

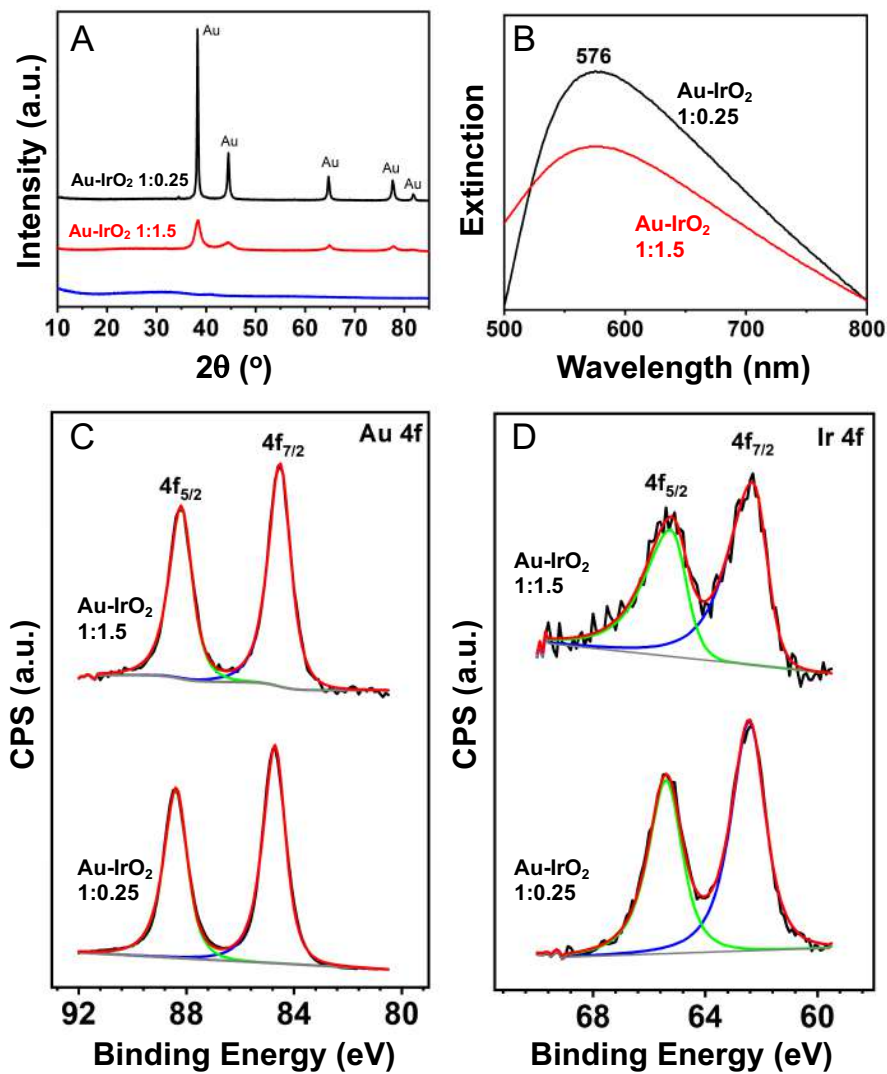

**Figure 3.** XRD diffractograms (A) and UV-VIS extinction spectra (B) recorded for Au-$IrO_2$ nanodendrites obtained by co-reduction of $AuCl_4^-$ and $IrCl_3 \cdot xH_2O$ as precursors in a 1:1.5 (red trace) 1:0.25 (black trace) molar ratios. (C) and (D) show the deconvoluted X-ray photoelectron spectra (XPS) of the Au 4f (C) and Ir 4f (D) core levels for the Au-$IrO_2$ nanodendrites obtained by co-reduction of $AuCl_4^-$ and $IrCl_3 \cdot xH_2O$ as precursors in a 1:1.5 (top trace) and 1:0.25 (bottom trace) molar ratios.

To study the surface composition, we also characterized the samples by XPS. **Figures 3C** and **D** depict the photoemission spectra in the Au 4f (**Figure 3C**) and Ir 4f (**Figure 3D**) core level regions for the nanodendrites prepared employing $AuCl_4^-$:$IrCl_3 \cdot xH_2O$ 1:1.5 (top trace) and 1:0.25 (bottom trace) molar ratios. A summary of the binding energy (BE) values and calculated surface composition is shown in **Table 1**. The Au 4f region showed two intense photoelectron peaks with maxima at BE of 84 and 88 eV, ascribed to Au $4f_{7/2}$ and Au $4f_{5/2}$, respectively. These values are consistent with the presence of Au species in the metallic state.[65] However, a slight negative shift from 88.4 and 84.7 eV to 88.2 and 84.5 eV in the samples prepared under $AuCl_4^-$:$IrCl_3 \cdot xH_2O$ 1:0.25 and 1:1.5 molar ratios, respectively, was observed. These variations can be ascribed to intrinsic charge transfer between $IrO_2$ and Au, which was more pronounced when the amount of $IrO_2$ in the material was higher (1:1.5 sample). In the Ir 4f core level region, the XPS spectra displayed two photoelectron emission peaks at binding energies of around 62 and 65 eV (**Figure 3D**, **Table 1**), corresponding to the Ir $4f_{7/2}$ and $4f_{5/2}$ states, respectively. The peak position and the asymmetry of the peaks evidences the presence of oxidized Ir species ($Ir^{4+}$) consistent with the formation of $IrO_2$ since the BE values reported for metallic Ir and $IrO_2$ standards are 61.1/64.1 eV and 62.7/65.7 eV, respectively.[66,67] The detected shifts in the Ir peak position in the two samples were very small, and both values are close to what is observed in IrO2 materials.[66,67] This is important since STEM-EDS elemental mapping could not confirm whether the Ir was present as metallic Ir or $IrO_2$ due to oxygen signal being overwhelmed by the presence of oxygen containing surface adsorbates. Finally, the surface composition estimated from the XPS spectra and Ir/Au ratio (**Table 1**) revealed an increase in the Ir content at the surface as the amount $IrCl_3 \cdot xH_2O$ precursor relative to $AuCl_4^-$ was increased during the synthesis. The Ir contents in **Table 1** are both higher than the from STEM-EDX and FAAS, further demonstrating that the Ir exists as a surface layer.

**Table 1:** Binding energies (eV) values and surface composition measured by XPS.

| Au-IrO$_2$ / AuCl$_4^-$:IrCl$_3 \cdot$xH$_2$O molar ratios | Au 4f (eV) | | Ir 4f (eV) | | Surface composition | | |
|---|---|---|---|---|---|---|---|
| | 4f$_{5/2}$ | 4f$_{7/2}$ | 4f$_{5/2}$ | 4f$_{7/2}$ | Au(%) | Ir(%) | Ir/Au |
| 1:0.25 | 88.4 | 84.7 | 65.4 | 62.4 | 92.0 | 8.0 | 0.09 |
| 1:1.5 | 88.2 | 84.5 | 65.2 | 62.3 | 80.5 | 19.5 | 0.24 |

In the synthesis of the nanodendrites shown in **Figures 1** and **2**, the Au and Ir precursors are firstly mixed at room temperature, which is accompanied by a change in the color of the solution from to green to blue. This change in color indicates the reduction of AuCl$_4^-$ species to AuCl$_2^-$ by IrCl$_6^{3-}$ (leading to the formation of Ir$^{4+}$ species).[56] In this case, we postulate that, as AuCl$_2^-$ has a higher standard reduction potential relative to AuCl$_4^-$, this species is mainly responsible for the formation of the Au nuclei during the synthesis. Then, as further Au and Ir are produced from the reduction of precursors (co-reduction), growth takes place by precursor addition at the surface of the pre-formed nuclei/seeds as well as by oriented attachment in the presence of citrate, which produces the nanodendrite morphology. It is important to note that the oriented attachment mechanism has been reported during the formation of Au nanostructures in the presence of citrate, and has been recently been confirmed by liquid cell TEM studies.[68,69]

We performed a series of experiments in order to confirm this proposed mechanism. We started by monitoring the optical properties of the precursor solutions before and after they are mixed at room temperature. **Figure S7A** shows the absorption spectra for the AuCl$_4^-$ and IrCl$_3 \cdot$xH$_2$O precursor solutions (red and blue traces, respectively). While AuCl$_4^-$ has an absorption band at 310 nm, only an intense signal below 290 nm was detected for the IrCl$_3 \cdot$xH$_2$O solution. Therefore, we employed the band at 310 nm to monitor the AuCl$_4^-$ reduction when the AuCl$_4^-$ and IrCl$_3$•xH$_2$O precursors were mixed. **Figures S7B** and **C** show the UV-VIS spectra for the mixtures of the precursor solutions in the AuCl$_4^-$:IrCl$_3$•xH$_2$O 1:1.5 and 1:0.25 molar ratios, respectively (which corresponded to the synthesis conditions described in **Figure 1** and **2**). In each case, the UV-VIS spectra were recorded at 5 s intervals following the mixture of the precursors. It can be observed that, in both cases, as the precursors were mixed, a gradual disappearance of the band assigned to AuCl$_4^-$ was detected. This agrees with the AuCl$_4^-$ being reduced to AuCl$_2^-$.[70,71] No LSPR peaks assigned to Au nanoparticles were observed, indicating that no further reduction to Au$^0$ takes place simply by mixing the precursors. Interestingly, under otherwise identical conditions, the AuCl$_4^-$ reduction kinetics were strongly dependent on the amount of IrCl$_3$•xH$_2$O precursor employed during the synthesis. The AuCl$_4^-$ reduction was much faster with a higher IrCl$_3$•xH$_2$O content and we postulate that this increased reduction kinetics favored in an increase in the number of Au nuclei, and hence a reduction in the overall size of the nanodendrites. In support of this hypothesis, we found it was possible to fine tune the size of the nanodendrites by controlling the AuCl$_4^-$:IrCl$_3$•xH$_2$O molar ratio during synthesis. Experiments employing AuCl$_4^-$:IrCl$_3$•xH$_2$O molar ratios of 1:1.25, 1:1, 1:0.75, and 1:0.5 led to nanodendrites having diameters of 98, 100, 115, and 130 nm, respectively (**Figure S8**).

We also studied the kinetics of the reaction by studying the nanodentrites formed at different time intervals following the addition of the solution containing the mixture of Ir and Au precursors to the boiling citrate solution. This was performed by removing aliquots from the reaction mixture, immersing in an ice bath, and isolating/washing the nanostructures by successive rounds of centrifugation and removal of the supernatant. SEM images for the Au-IrO$_2$ nanodendrites extracted from the reaction mixture after 15 s, 1 min, 5 min, 10 min, and 20 min are shown in **Figure S9**. The dendritic morphology can already be observed in the products obtained after 15 s of synthesis. Moreover, the SEM images show that the nanodendrites gradually increase in size as a function of time, which agrees with a mechanism based on oriented attachment.[72] It is important to note that the nanodendrite morphology was not observed when the synthesis was performed in the presence of only $AuCl_4^-$ or $IrCl_3 \cdot xH_2O$ precursors (**Figure S10**). In this case, spherical Au particles or irregular Ir-based materials were obtained. This result indicates that the presence of both precursors is required for formation of the nanodendritic morphology, in agreement with the proposed mechanism based on the formation of $AuCl_2^-$, by $IrCl_3 \cdot xH_2O$, which then reduces to Au nuclei during synthesis.

It is also interesting to investigate how the choice of citrate as a stabilizer affects the nanodendrite morphology. Oriented attachment in Au nanocrystals stabilized by citrate is reported to occur with attachment observed preferentially for {111} orientations[68,69] Images of the products obtained under similar conditions but replacing citrate by PVP, hydroquinone, ascorbic acid, or borohydride led to the formation of irregular particles (**Figure S11**). This indicates that citrate, in addition to being a reducing and capping agent, is also an essential component for formation of the nanodentrite morphology, playing a key role in the aggregation and oriented attachment processes.

To better understand the how the nanodendrite morphology can affect the plasmonic properties of the Au we have performed simulations for the electric field distribution and the extinction, absorption, and scattering spectra in the 400-700 nm range using the discrete dipole approximation (DDA) method. Our DDA simulations show that the regions between individual Au branches (modelled as a dimer or close packing of spherical Au NPs) can display much higher electric field enhancement as well higher extinction and absorption efficiencies relative to isolated particles (**Figure S1**).

Finally, the Au-IrO$_2$ nanodendrites were employed as model systems to investigate how plasmonic effects in Au can be harnessed towards the enhancement of the electrocatalytic activity of IrO$_2$ towards the OER. The plasmonic effects over electrocatalytic performance were studied in O$_2$-saturated 0.1 M KOH electrolyte at room temperature via a typical three-electrode system at a scan rate of 10 mVs$^{-1}$ at 532 nm irradiation (200 mW). **Figure 4A** shows linear sweep voltammetry (LSV) tests for the nanodendrites obtained with 1:1.5 and 1:0.25 AuCl$_4^-$:IrCl$_3 \bullet$xH$_2$O molar ratios in the presence of light excitation (solid lines) and in the absence of light (dashed lines) (CVs are shown in **Figure S12**). The measured current for the bare glassy carbon electrode is also shown for comparison (black trace). It can be observed that the Au-IrO$_2$ 1:1.5 sample displayed higher current densities and an earlier onset potential relative to Au-IrO$_2$ 1:0.25, which can be ascribed to its superior OER activity associated with the higher Ir loading in this material. Importantly, when irradiated at 532 nm, the OER is accelerated in both samples. Here, a significant decrease in onset potential and increase in the detected current densities was detected under visible light illumination, assigned to the plasmonic enhancement of the OER.[40] Specifically, the detected potentials required to achieve a current density of 10 mA cm$^{-2}$, which represent an important metric in the solar synthesis of fuels, decreased from 0.72 to 0.69 for the Au-IrO$_2$ 1:1.5 sample under visible light illumination. For Au-IrO$_2$ 1:0.25, a decrease from 0.85 to 0.78 V took place. Considering the E$_{OER}$ value of 0.404 V in alkaline solution, the calculated overpotential at 10 mA cm$^{-2}$ for Au-IrO$_2$ 1:1.5 and Au-IrO$_2$ 1:0.25 samples corresponded to 286 and 376 mV, respectively. **Table 2** summarises the reported record values of overpotential for IrO$_2$ and RuO$_2$ electrocatalysts, which are typically in the 300-400 mV range. Therefore, an over potential of just 286 mV for the plasmonically assisted OER activity in the Au-IrO$_2$ 1:1.5 molar ratio represents a record value for IrO$_2$ electrocatalysts which matches the most active reported electrocatalysts.[32, 40, 56, 73, 74, 75, 76, 77, 78]

In order to provide more insights into the effect of LSPR excitation on the kinetics of the OER, an analysis of the Tafel slope was performed. As seen from **Figure 4B**, the Au-IrO$_2$ 1:1.5 sample exhibited a Tafel slope of 120 mV dec$^{-1}$ in the dark (dashed red line). Upon plasmonic excitation, this value is sharply decreased to 76 mV dec$^{-1}$ (solid red line), clearly showing that the kinetics of water oxidation are accelerated by visible light excitation.[79] For the Au-IrO$_2$ 1:0.25 sample, a similar decrease in the Tafel slope value from 130 to 100 mV dec$^{-1}$ was detected upon visible light illumination (dashed and solid blue lines, respectively).

In order to demonstrate the effect of the LSPR excitation over the electrocatalytic activity, we also collected the *I-t* curve at 0.65 V (*vs* Ag/AgCl) for Au-IrO$_2$ 1:1.5 (**Figure 4C**) and Au-IrO$_2$ 1:0.25 (**Figure 4D**) samples under chopped light illumination for 3 different excitation wavelengths: 405 (blue trace), 523 (green trace), and 638 nm (red trace). From **Figures 4C** and **D** it can be seen that the Au-IrO$_2$ samples displayed fast and reproducible current responses to on-off illumination cycles. Moreover, the current densities were clearly wavelength dependent, being greatest for 532 nm excitation, followed by 405 and 638 nm. Therefore, the current density enhancement over light illumination matched the extinction spectra of the Au-IrO$_2$ nanodendrites, being strongest when the overlap between the excitation wavelength and the position of the LSPR band was the highest. **Figure 4C** and **D** also indicate that there was a decrease in the detected currents as a function of testing time under light irradiation. This behavior indicates loss of OER performance (**Figure S13**), indicating that stability of the Au-IrO$_2$ nanostructures needs further optimizations.

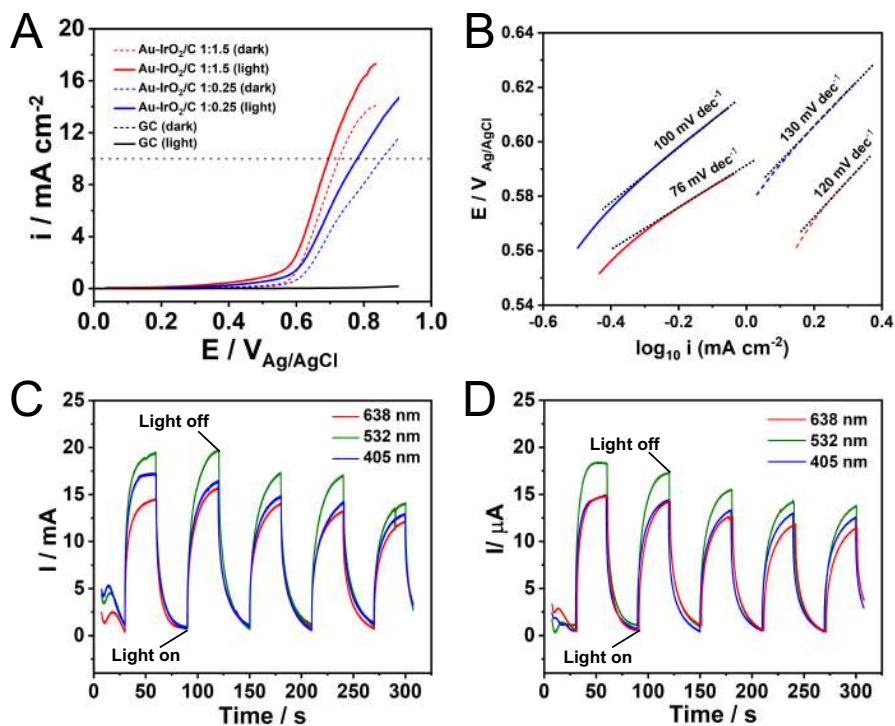

**Figure 4.** Electrochemical performances of the Au-IrO$_2$ materials measured in a 0.1 M KOH electrolyte with and without SPR excitation. OER linear scanning voltammetries recorded at a scan rate = 10 mVs$^{-1}$ (A) and Tafel plots (B) for the Au-IrO$_2$ nanodendrites obtained by co-reduction employing AuCl$_4^-$ and IrCl$_3$·xH$_2$0 as precursors in a 1:1.5 (red trace) 1:0.25 (blue trace) molar ratios with light excitation at 532 nm (solid line) and without light excitation (dashed line) . The dotted horizontal line in (A) indicates a 10 mAcm$^{-1}$ current. The measured current for the bare glassy carbon electrode is also shown for comparison (black trace).  (C) and (D) show the light wavelength I–t curves recorded at 0.65 V (vs Ag/AgCl) with light irradiation on/off for Au-IrO$_2$ nanodendrites obtained by co-reduction employing AuCl$_4^-$ and IrCl$_3$·xH$_2$0 as precursors in a 1:1.5 (C) 1:0.25 (D) molar ratios under 405 (blue trace), 532 (green trace), and 638 nm (red trace) excitation. All the experiments were performed at room temperature.

Table 2. Comparison on the OER Activity, expressed by the values for the overpotential (η) at j = 10 mAcm$^{-2}$ for various catalysts (plasmonic and nonplasmonic).

| Catalyst | η at j = 10 mA cm$^{-2}$ (mV) | Ref |
|---|---|---|
| Au-IrO$_2$ 1:1.5 (light) | 286 | This work |
| Au-IrO$_2$ 1:1.5 (dark) | 326 | This work |
| Au-IrO$_2$ 1:0.25 (light) | 376 | This work |
| Au-IrO$_2$ 1:0.25 (dark) | 446 | This work |
| IrOx/Au | 370 | 80 |
| IrOx[0.05]-Au nanoflowers | 481 | 56 |
| IrO$_2$ | 330 | 73 |
| RuO$_2$ | 305 | 74 |
| Au (light) | 455 | 40 |
| Au (dark) | 573 | 40 |
| CoFe$_2$O$_4$ | 370 | 75 |
| CaFeO$_3$ | 390 | 76 |
| g-Ni$_{0.87}$Fe$_{0.13}$OOH | 390 | 32 |
| NiFe DH | 290 | 77 |
| Ni(OH)$_2$–Au (light) | 270 | 40 |
| Ni(OH)$_2$–Au (dark) | 330 | 40 |
| AuNP@Co/Ni-MOF | 330 | 78 |

It is recognized that the electrocatalyzed OER is a heterogeneous reaction comprising multiple elementary steps involving four electron transfer processes (multiple reaction channels) and generating several intermediates, such as OH*, O* and OOH*.[32] Most of the proposed mechanisms include the formation of intermediates such as MOH and MO.[79,81,82] Under basic conditions, the mechanism proceeds as described in **Equations 1-5**.[32,79,81–83] In these processes, the bonding interactions (M–O) within the intermediates (MOH, MO and MOOH) are crucial for the overall electrocatalytic process.[32,79,81–83]

$$M + OH^- \rightarrow MOH + e^- \qquad \text{Eq. 1}$$

$$MOH + OH^- \rightarrow MO + H_2O_{(l)} + e^- \qquad \text{Eq. 2}$$

$$MO + OH^- \rightarrow MOOH + e^- \qquad \text{Eq. 3}$$

$$MOOH + OH^- \rightarrow MOO + H_2O_{(l)} + e^- \qquad \text{Eq. 4}$$

$$MOO \rightarrow M + O_{2(g)} \qquad \text{Eq. 5}$$

We performed DFT calculations in order to understand the reason that LSPR excitation produces increased OER activities for our photocatalytic materials. We calculated the binding energies and electronic structure of the O* and OH* intermediates for two slab models: *i*) two IrO$_2$ (110) layers (**Figure S14**), the thickness of which is ~ 1 nm, matching the experimental observation (**Figure 1**) and *ii*) two IrO$_2$ (110) layers supported on three Au (111) layers to simulate the Au-IrO$_2$ material. **Figure 5A** shows the two IrO$_2$ (110) layers supported on three Au (111) layers employed in our model and **Figure 5B** shows the O species adsorbed at the IrO$_2$ surface. The calculated O* and OH* binding energies at the IrO$_2$ and IrO$_2$/Au surfaces are shown in **Figure**

**5C**. We find that adsorption of both the O* and OH* is enhanced on the IrO$_2$/Au surfaces relative to clean IrO$_2$ surface. Note it has been recently demonstrated that the ΔG(O*)-ΔG(OH*) can be employed as a descriptor of OER activity, in which a volcano plot relationship between ΔG(O*)-ΔG(OH*) values and OER activity has been established.[84,85] Our DFT calculations showed that ΔG(O*)-ΔG(OH*) for IrO$_2$/Au was higher than for IrO$_2$; 1.30 and 1.19 eV for IrO$_2$/Au and IrO$_2$ respectively (**Figure 5C**). The increase in the value of ΔG(O*)-ΔG(OH*) brings it closer to the maximum of the volcano plot, indicating that the OER activity for Au-IrO$_2$ should be higher than it on IrO$_2$.[85]

**Figure 5D** shows the calculated projected density of states (DOS) for O* and OH* adsorbed on Au-IrO$_2$ (**Figure S15** shows the calculated DOS for O* and OH* adsorbed on IrO$_2$ for comparison). It was found that the states between -2.33 eV (corresponding to the photon wavelength of 532 nm used in the experiments) and the Fermi level are more pronounced than the states above the Fermi level. This high population of states below the Fermi level may indicate a more pronounced mechanism based on the effect of the hot holes generated under LSPR excitation towards the enhanced OER activities.[86,87] In **Figure 5E**, the charge density analysis of an energy range from -2.33 to the Fermi level for O* and OH* on IrO$_2$/Au shows the presence of more states on *O, which agrees with the DOS plot (**Figure 5D**). Based on these results, we propose that when hot holes are LSPR-excited in the Au NPs and then transferred to O* (or the Ir-O antibonding state), the charge lowers the binding energy of O* increasing the value of ΔG(O*), and thus increases the OER activity. It is plausible that the hot holes may also lower the binding energy of OH*. However, because OH* has much less available states to populate the holes (**Figure 5D**), this weakening of the binding energy of OH* is much less than the O* binding, which overall makes ΔG(O*)-ΔG(OH*) more positive under LSPR excitation, further shifting this value closer towards the maximum point in the volcano plot between ΔG(O*)-ΔG(OH*) and the OER activity.[85]

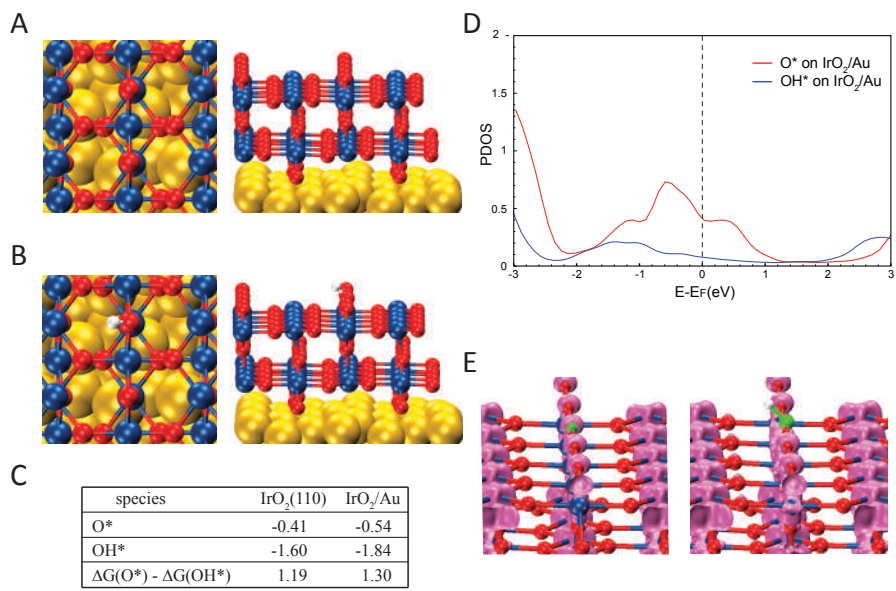

**Figure 5.** (A) The slab models employed in the DFT calculations consisting of two $IrO_2$ (110) on three Au (111) layers. (B) Calculated atomic structure of the O species adsorbed on the $IrO_2$ (110) surface in the Au-$IrO_2$ hybrid structure. (C) Calculated binding energies of O* and OH* species and ΔG(O*)-ΔG(OH*) values on Au-$IrO_2$ and the pure $IrO_2$ slab models. When Au is present as the support for $IrO_2$, the ΔG(O*)-ΔG(OH*) values (1.30 vs 1.19 eV) is closer to the optimal point on the volcano plot for maximizing the OER activity. (D) Projected density of states of O and OH adsorbed on Au-$IrO_2$. (E) Charge density analysis of energy range (-2.33,0 eV) for O* and OH* on Au-$IrO_2$, showing more occupied states on O*.

Our DFT calculations agree with the experimentally observed decrease in the Tafel slopes under visible light illumination, which indicated that the OER kinetics of the Au-$IrO_2$ nanodendrites can be facilitated by LSPR excitation, in which hot holes can be generated at the Au NPs and flow to the $IrO_2$ surface, where enhance the OER process. As the Tafel slopes depend on the strength of adsorption of the intermediate species, this would support our proposed mechanism in which hot holes can activate the M-O* bond at the surface during the OER, lowering its binding energy, which leads to a change in the rate limiting step.[32,35,79,81,82] It has been reported that the OER reaction mechanism and rate-determining step cannot be unambiguously assigned simply according to the measured Tafel slope because surface intermediates, coverage, reaction pathways, and rate-determining steps may depend on the potential.[32,35,79,81,82] Nevertheless, it has been proposed that if the first-electron transfer represents the rate-determining step, the corresponding Tafel slope should correspond to 120 mVdec$^{-1}$. If the rate-determining step is the chemical reaction following a one-electron transfer process, the Tafel slope becomes 60 mVdec$^{-1}$.[32,35,79,81,82] One example is a process in which an OH surface species is rearranged via a surface reaction, as described in **Eq. 4.** Therefore, our experimentally observed change in the Tafel slope from 120 to 76 mVdec$^{-1}$ in the Au-IrO$_2$ material as a result of visible light illumination may indicate that the OER mechanism and thus the rate determining step is changing from the first electron transfer reaction (**Eq. 1**) to the chemical reaction (**Eq. 4**).

It is important to note that, although we focused herein on the OER activities in alkaline conditions, we believe that the established the design principles can also be applied for developing plasmonic-catalytic electrocatalysts for the OER in neutral and acidic media. Though the adsorption energies of O and OH were found to shift to more positive energies with lower pH, it has been observed that the free energy shifts for OH* and O* preserves the universal scaling relation between free energy difference of O* and OH* with OER activity.[88]

Also, the transfer of energy from the LSPR excitation to the reactant for the acceleration of the OER can occur via LSPR generated hot charges or an electronic excitation.[8] This process does not require charge extraction from the metal, being able to take place by a transient electronic exchange between the metal and reactant. This leads to the formation of transient adsorbate can induce chemical transformations or add vibrational energy to the reactant facilitating the reaction.[8] In the case of charge extraction from the metal, LSPR excited hot electrons be transferred to the external circuit or to electron accepting adsorbates.[89] In addition, the electric field enhancements at the surface of the Au-IrO$_2$ nanostructures can play an important role in the activation of Ir-O bonds in the case of a direct transfer based on electronic excitation of metal-molecule interfacial states and as their energy can be dissipated through the nonradiative excitation of energetic charge carriers.[58,90] Finally, temperature effects (localized heating following LSPR decay ) may also play a role (together with LSPR hot carriers) over the OER activity enhancements. While the enhancement of electrocatalytic activity can originate from both photothermal effects and LSPR-generated charge carriers, previous results have shown that photothermally induced temperature rise does not fully account for the enhanced electrocatalytic activity, and that non-thermal effect play a significant role. [89]

**Conclusion**

We have developed multimetallic Au-IrO$_2$ plasmonic-catalytic nanoparticles and successfully demonstrated enhancement of the OER activity of IrO$_2$ via plasmonic catalysis under visible light illumination. The Au-IrO$_2$ plasmonic-catalytic nanoparticles we synthesized were tunable for a range of sizes and compositions but all comprised of a core-shell Au-IrO$_2$ dendritic nanoparticle morphology with closely spaced Au dendrite branches partially covered by an ultrathin (1 nm) IrO$_2$ shell. These morphological and compositional features meet important design principles towards the optimization of OER activities and allow us to expand our understanding of enhancement mechanisms due to coupling with surface plasmon effects. The ultrathin and incomplete IrO$_2$ shell at the surface of each branch enabled one to maximize the light harvesting by Au, and the plasmonic hybridization between the closely spaced plasmonic branches allow for the generation of electromagnetic hot spots to enhance light interactions. The OER activities we measured were wavelength dependent, being maximized at wavelengths that matched the LSPR energies, with the best values equaling the most active catalysts reported for OER. Furthermore, our calculations suggest that light generated hot holes play a dominant role in the mechanism of plasmonic enhancement. These LSPR excited holes could be transferred to Ir-O antibonding states, lowering the binding energy and accelerating the reaction. This was further confirmed by Tafel plot analysis, which suggested a change in the reaction mechanism (rate-determinant step) under LSPR excitation. We believe the results reported herein shed novel insights into the design principles required to develop plasmonic-catalytic nanoparticles capable of optimizing activities and also enable further mechanistic understanding over enhancement mechanisms that dictate plasmon-driven chemistry.

**Acknowledgments**

This work was supported by FAPESP (Grant 2015/26308-7) and the Serrapilheira Institute (Grant Serra-1709-16900). This study was financed in part by the Coordenação de Aperfeiçoamento de Pessoal de Nível Superior – Brazil (CAPES) – Finance Code 001. L.S.P., J.Q. and E.C.M.B. thank FAPESP for the fellowships (Grants 2016/00819-8, 2016/17866-9, and 2015/11452-5, respectively). I.C.F. and B.A.N. thanks CNPq for the fellowship. T.M. and B.W. appreciate support from the Department of Energy (Grant No. DE-SC0020300) and computational resources at the OU Supercomputing Center for Education & Research (OSCER) at the University of Oklahoma. S.J.H. thanks the Engineering and Physical Sciences (U.K.) (Grants EP/M010619/1, EP/P009050/1) and the European Research Council (ERC) under the European Union's Horizon 2020 research and innovation programme (Grant ERC-2016-STG-EvoluTEM-715502).

**References**

1       Y. Zhang, S. He, W. Guo, Y. Hu, J. Huang, J. R. Mulcahy and W. D. Wei, *Chem. Rev.*, 2018, **118**, 2927–2954.

2       G. Baffou and R. Quidant, *Chem. Soc. Rev.*, 2014, **43**, 3898–3907.

3       S. Linic, U. Aslam, C. Boerigter and M. Morabito, *Nat. Mater.*, 2015, **14**, 567–576.

4       R. Long, Y. Li, L. Song and Y. Xiong, *Small*, 2015, **11**, 3873–3889.

5       J. G. Smith, J. A. Faucheaux and P. K. Jain, *Nano Today*, 2015, 10, 67–80.

6       S. Yu, A. J. Wilson, G. Kumari, X. Zhang and P. K. Jain, *ACS Energy Lett.*, 2017, **2**, 2058–2070.

7       M. L. Brongersma, N. J. Halas and P. Nordlander, *Nat. Nanotechnol.*, 2015, **10**, 25–34.

8       U. Aslam, V. G. Rao, S. Chavez and S. Linic, *Nat. Catal.*, 2018, **1**, 656–665.

9       M. J. Kale, T. Avanesian and P. Christopher, *ACS Catal.*, 2014, **4**, 116–128.

10      K. Li, N. J. Hogan, M. J. Kale, N. J. Halas, P. Nordlander and P. Christopher, *Nano Lett.*, 2017, **17**, 3710–3717.

11  J. Quiroz, E. C. M. Barbosa, T. P. Araujo, J. L. Fiorio, Y. C. Wang, Y. C. Zou, T. Mou, T. V. Alves, D. C. De Oliveira, B. Wang, S. J. Haigh, L. M. Rossi and P. H. C. Camargo, *Nano Lett.*, 2018, **18**, 7289–7297.

12  D. F. Swearer, H. Zhao, L. Zhou, C. Zhang, H. Robatjazi, J. M. P. Martirez, C. M. Krauter, S. Yazdi, M. J. McClain, E. Ringe, E. A. Carter, P. Nordlander and N. J. Halas, *Proc. Natl. Acad. Sci.*, 2016, **113**, 8916–8920.

13  A. Marimuthu, J. Zhang and S. Linic, *Science*, 2013, **340**, 1590–1593.

14  C. Hu, X. Chen, J. Jin, Y. Han, S. Chen, H. Ju, J. Cai, Y. Qiu, C. Gao, C. Wang, Z. Qi, R. Long, L. Song, Z. Liu and Y. Xiong, *J. Am. Chem. Soc.*, 2019, **141**, 7807–7814.

15  E. Peiris, S. Sarina, E. R. Waclawik, G. A. Ayoko, P. Han, J. Jia and H. Y. Zhu, *Angew. Chem. Int. Ed.*, 2019, **58**, 12032–12036.

16  J. L. Wang, R. A. Ando and P. H. C. Camargo, *ACS Catal.*, 2014, **4**, 3815–3819.

17  A. G. M. A. G. M. A. G. M. Da Silva, T. S. T. S. Rodrigues, J. Wang, L. K. L. K. Yamada, T. V. T. V. Alves, F. R. Ornellas, R. A. R. A. Ando and P. H. C. P. H. C. Camargo, *Langmuir*, 2015, **31**, 10272–10278.

18  S. Linic, P. Christopher and D. B. Ingram, *Nat. Mater.*, 2011, **10**, 911–921.

19  S. Yu and P. K. Jain, *Nat. Commun.*, 2019, **10**, 2022.

20  Y. Kim, J. G. Smith and P. K. Jain, *Nat. Chem.*, 2018, **10**, 763–769.

21  M. Dhiman, A. Maity, A. Das, R. Belgamwar, B. Chalke, Y. Lee, K. Sim, J.-M. Nam and V. Polshettiwar, *Chem. Sci.*, 2019, **10**, 6594–6603.

22  S. Atta, A. M. Pennington, F. E. Celik and L. Fabris, *Chem*, 2018, **4**, 2140–2153.

23  S. Mukherjee, F. Libisch, N. Large, O. Neumann, L. V. Brown, J. Cheng, J. B. Lassiter, E. A.

Carter, P. Nordlander and N. J. Halas, *Nano Lett.*, 2012, **13**, 240–247.

24      S. Mukherjee, L. Zhou, A. M. Goodman, N. Large, C. Ayala-Orozco, Y. Zhang, P. Nordlander and N. J. Halas, *J. Am. Chem. Soc.*, 2014, **136**, 64–67.

25      P. Christopher, H. Xin and S. Linic, *Nat. Chem.*, 2011, **3**, 467–472.

26      M. J. Landry, A. Gellé, B. Y. Meng, C. J. Barrett and A. Moores, *ACS Catal.*, 2017, **7**, 6128–6133.

27      P. Han, W. Martens, E. R. Waclawik, S. Sarina and H. Zhu, *Part. Part. Syst. Charact.*, 2018, 35, 1700489.

28      M. G. Walter, E. L. Warren, J. R. McKone, S. W. Boettcher, Q. Mi, E. A. Santori and N. S. Lewis, *Chem. Rev.*, 2010, **110**, 6446–6473.

29      N. S. Lewis and D. G. Nocera, *Proc. Natl. Acad. Sci.*, 2006, **103**, 15729–15735.

30      M. Tahir, L. Pan, F. Idrees, X. Zhang, L. Wang, J. J. Zou and Z. L. Wang, *Nano Energy*, 2017, 37, 136–157.

31      S. Bai, X. Li, Q. Kong, R. Long, C. Wang, J. Jiang and Y. Xiong, *Adv. Mater.*, 2015, **27**, 3444–3452.

32      N.-T. Suen, S.-F. Hung, Q. Quan, N. Zhang, Y.-J. Xu and H. M. Chen, *Chem. Soc. Rev.*, 2017, **46**, 337–365.

33      H. Dotan, K. Sivula, M. Grätzel, A. Rothschild and S. C. Warren, *Energy Environ. Sci.*, 2011, **4**, 958–964.

34      S. Trasatti, *J. Electroanal. Chem.*, 1980, **111**, 125–131.

35      E. Fabbri, A. Habereder, K. Waltar, R. Kötz and T. J. Schmidt, *Catal. Sci. Technol.*, 2014, 4, 3800–3821.

36   P. G. Hoertz, Y. Il Kim, W. J. Youngblood and T. E. Mallouk, *J. Phys. Chem. B*, 2007, **111**, 6845–6856.

37   Y. Zhao, E. A. Hernandez-Pagan, N. M. Vargas-Barbosa, J. L. Dysart and T. E. Mallouk, *J. Phys. Chem. Lett.*, 2011, **2**, 402–406.

38   B. S. Yeo and A. T. Bell, *J. Am. Chem. Soc.*, 2011, **133**, 5587–5593.

39   B. S. Yeo and A. T. Bell, *J. Phys. Chem. C*, 2012, **116**, 8394–8400.

40   G. Liu, P. Li, G. Zhao, X. Wang, J. Kong, H. Liu, H. Zhang, K. Chang, X. Meng, T. Kako and J. Ye, *J. Am. Chem. Soc.*, 2016, **138**, 9128–9136.

41   L. Guo, K. Liang, K. Marcus, Z. Li, L. Zhou, P. D. Mani, H. Chen, C. Shen, Y. Dong, L. Zhai, K. R. Coffey, N. Orlovskaya, Y.-H. Sohn and Y. Yang, *ACS Appl. Mater. Interfaces*, 2016, **8**, 34970–34977.

42   F. Shi, J. He, B. Zhang, J. Peng, Y. Ma, W. Chen, F. Li, Y. Qin, Y. Liu, W. Shang, P. Tao, C. Song, T. Deng, X. Qian, J. Ye and J. Wu, *Nano Lett.*, 2019, **19**, 1371–1378.

43   H.-X. Zhang, Y. Li, M.-Y. Li, H. Zhang and J. Zhang, *Nanoscale*, 2018, **10**, 2236–2241.

44   X. Guo, X. Li, S. Kou, X. Yang, X. Hu, D. Ling and J. Yang, *J. Mater. Chem. A*, 2018, **6**, 7364–7369.

45   J. E. Lee, F. Marques Mota, C. H. Choi, Y. R. Lu, R. Boppella, C. L. Dong, R. S. Liu and D. H. Kim, *Adv. Mater. Interfaces*, 2019, **6**, 1801144.

46   J. H. Weaver, *Phys. Rev. B*, , DOI:10.1103/PhysRevB.11.1416.

47   D. F. Swearer, H. Robatjazi, J. M. P. Martirez, M. Zhang, L. Zhou, E. A. Carter, P. Nordlander and N. J. Halas, *ACS Nano*, 2019, **13**, 8076–8086.

48   H. Robatjazi, H. Zhao, D. F. Swearer, N. J. Hogan, L. Zhou, A. Alabastri, M. J. McClain, P.

Nordlander and N. J. Halas, *Nat. Commun.*, 2017, **8**, 27.

49  H. Liu, X. Meng, T. D. Dao, L. Liu, P. Li, G. Zhao, T. Nagao, L. Yang and J. Ye, *J. Mater. Chem. A*, 2017, **5**, 10567–10573.

50  Z. Li, L. Shi, D. Franklin, S. Koul, A. Kushima and Y. Yang, *Nano Energy*, 2018, **51**, 400–407.

51  U. Aslam, S. Chavez and S. Linic, *Nat. Nanotechnol.*, 2017, **12**, 1000.

52  T. P. Araujo, J. Quiroz, E. C. M. Barbosa and P. H. C. Camargo, *Curr. Opin. Colloid Interface Sci.*, 2019, 39, 110–122.

53  E. Cortés, *Adv. Opt. Mater.*, 2017, **5**, 1700191.

54  C. C. L. McCrory, S. Jung, J. C. Peters and T. F. Jaramillo, *J. Am. Chem. Soc.*, 2013, **135**, 16977–16987.

55  T. Reier, M. Oezaslan and P. Strasser, *ACS Catal.*, 2012, **2**, 1765–1772.

56  C. Zhao, Y. E. and L. Fan, *Microchim. Acta*, 2012, **178**, 107–114.

57  C. Zhang, H. Zhao, L. Zhou, A. E. Schlather, L. Dong, M. J. McClain, D. F. Swearer, P. Nordlander and N. J. Halas, *Nano Lett.*, 2016, **16**, 6677–6682.

58  S. Chavez, U. Aslam and S. Linic, *ACS Energy Lett.*, 2018, **3**, 1590–1596.

59  L. Zhou, D. F. Swearer, C. Zhang, H. Robatjazi, H. Zhao, L. Henderson, L. Dong, P. Christopher, E. A. Carter, P. Nordlander and N. J. Halas, *Science*, 2018, **362**, 69–72.

60  F. Wang, C. Li, H. Chen, R. Jiang, L.-D. Sun, Q. Li, J. Wang, J. C. Yu and C.-H. Yan, *J. Am. Chem. Soc.*, 2013, **135**, 5588–5601.

61  P. Christopher, H. Xin, A. Marimuthu and S. Linic, *Nat. Mater.*, 2012, **11**, 1044–1050.

62  S. Linic and M. A. Barteau, *J. Am. Chem. Soc.*, 2003, **125**, 4034–5.

63  A. G. M. da Silva, T. S. Rodrigues, V. G. Correia, T. V Alves, R. S. Alves, R. A. Ando, F. R.

Ornellas, J. Wang, L. H. Andrade and P. H. C. Camargo, *Angew. Chem. Int. Ed.*, 2016, **55**, 7111–7115.

64  A. G. M. da Silva, T. S. Rodrigues, L. S. K. Taguchi, H. V. Fajardo, R. Balzer, L. F. D. Probst and P. H. C. Camargo, *J. Mater. Sci.*, 2015, **51**, 603–614.

65  M. P. Seah, G. C. Smith and M. T. Anthony, *Surf. Interface Anal.*, 1990, **15**, 293–308.

66  R. G. Haverkamp, A. T. Marshall and B. C. C. Cowie, *Surf. Interface Anal.*, 2011, **43**, 847–855.

67  S. J. Freakley, J. Ruiz-Esquius and D. J. Morgan, *Surf. Interface Anal.*, 2017, **49**, 794–799.

68  B. Jin, M. L. Sushko, Z. Liu, C. Jin and R. Tang, *Nano Lett.*, 2018, **18**, 6551–6556.

69  C. Zhu, S. Liang, E. Song, Y. Zhou, W. Wang, F. Shan, Y. Shi, C. Hao, K. Yin, T. Zhang, J. Liu, H. Zheng and L. Sun, *Nat. Commun.*, 2018, **9**, 1–7.

70  X. Zhao, Q. Wang, X. Zhang, Y. I. Lee and H. G. Liu, *Phys. Chem. Chem. Phys.*, 2016, **18**, 1945–1952.

71  M. D. Urović, R. Puchta, Ž. D. Bugarčić and R. Van Eldik, *Dalt. Trans.*, 2014, **43**, 8620–8632.

72  H. You and J. Fang, *Nano Today*, 2016, **11**, 145–167.

73  J. Huang, J. Chen, T. Yao, J. He, S. Jiang, Z. Sun, Q. Liu, W. Cheng, F. Hu, Y. Jiang, Z. Pan and S. Wei, *Angew. Chem. Int. Ed.*, 2015, **54**, 8722–8727.

74  P. Chen, K. Xu, T. Zhou, Y. Tong, J. Wu, H. Cheng, X. Lu, H. Ding, C. Wu and Y. Xie, *Angew. Chemie - Int. Ed.*, 2016, **55**, 2488–2492.

75  M. Li, Y. Xiong, X. Liu, X. Bo, Y. Zhang, C. Han and L. Guo, *Nanoscale*, 2015, **7**, 8920–8930.

76  S. Yagi, I. Yamada, H. Tsukasaki, A. Seno, M. Murakami, H. Fujii, H. Chen, N. Umezawa, H.

Abe, N. Nishiyama and S. Mori, *Nat. Commun.*, 2015, **6**, 8249.

77  O. Diaz-Morales, I. Ledezma-Yanez, M. T. M. Koper and F. Calle-Vallejo, *ACS Catal.*, 2015, **5**, 5380–5387.

78  M. Wang, P. Wang, C. Li, H. Li and Y. Jin, *ACS Appl. Mater. Interfaces*, 2018, **10**, 37095–37102.

79  T. Shinagawa, A. T. Garcia-Esparza and K. Takanabe, *Sci. Rep.*, 2015, **5**, 13801.

80  S. Moon, Y. Bin Cho, A. Yu, M. H. Kim, C. Lee and Y. Lee, *ACS Appl. Mater. Interfaces*, 2019, **11**, 1979–1987.

81  D. A. García-Osorio, R. Jaimes, J. Vazquez-Arenas, R. H. Lara and J. Alvarez-Ramirez, *J. Electrochem. Soc.*, 2017, **164**, E3321–E3328.

82  Y.-H. Fang and Z.-P. Liu, *ACS Catal.*, 2014, **4**, 4364–4376.

83  E. Antolini, *ACS Catal.*, 2014, **4**, 1426–1440.

84  B. M. Tackett, W. Sheng, S. Kattel, S. Yao, B. Yan, K. A. Kuttiyiel, Q. Wu and J. G. Chen, *ACS Catal.*, 2018, **8**, 2615–2621.

85  I. C. Man, H. Y. Su, F. Calle-Vallejo, H. A. Hansen, J. I. Martínez, N. G. Inoglu, J. Kitchin, T. F. Jaramillo, J. K. Nørskov and J. Rossmeisl, *ChemCatChem*, 2011, **3**, 1159–1165.

86  A. E. Schlather, A. Manjavacas, A. Lauchner, V. S. Marangoni, C. J. DeSantis, P. Nordlander and N. J. Halas, *J. Phys. Chem. Lett.*, 2017, **8**, 2060–2067.

87  P. Christopher and M. Moskovits, *Annu. Rev. Phys. Chem.*, 2017, **68**, 379–398.

88  D.-Y. Kuo, J. K. Kawasaki, J. N. Nelson, J. Kloppenburg, G. Hautier, K. M. Shen, D. G. Schlom and J. Suntivich, *J. Am. Chem. Soc.*, 2017, **139**, 3473–3479.

89  L. Huang, J. Zou, J. Y. Ye, Z. Y. Zhou, Z. Lin, X. Kang, P. K. Jain and S. Chen, *Angew. Chem.*

*Int. Ed.*, 2019, **58**, 8794–8798.

90　B. Seemala, A. J. Therrien, M. Lou, K. Li, J. P. Finzel, J. Qi, P. Nordlander and P. Christopher, *ACS Energy Lett.*, 2019, **4**, 1803–1809.